\documentclass[aps,prd,preprintnumbers,nofootinbib,floatfix,11pt]{revtex4}

\usepackage{amsfonts,amsmath,amssymb,mathrsfs,graphicx,epsfig,color,amsthm} 
\usepackage{hyperref}
\usepackage{ulem}
\allowdisplaybreaks[3]
\newcommand{\be}{\begin{equation}}
\newcommand{\ee}{\end{equation}}
\newcommand{\ba}{\begin{eqnarray}}
\newcommand{\ea}{\end{eqnarray}}

\newcommand{\Cr}[1]{{{\cal O} \left( r^{#1} \right)}}
\newcommand{\Crs}[1]{o\left(r_0^{#1}\right)}
\newcommand{\Crz}[1]{{{\cal O} \left( r_0^{~#1} \right)}}
\newtheorem{proposition}{Proposition}
\newtheorem{lemma}{Lemma}

\definecolor{color1}{rgb}{0.3,0.3,0.6}

\begin{document}
\begin{flushright}
  YITP-23-49
    \end{flushright}

\title{Asymptotic behavior of null geodesics near future null infinity IV:  \\Null-access theorem for generic asymptotically flat spacetime}
\author{Masaya Amo$^{1,2}$, Keisuke Izumi$^{3,4}$, Yoshimune Tomikawa$^5$, Tetsuya Shiromizu$^{4,3}$ and Hirotaka Yoshino$^{6,7}$ }

\affiliation{$^{1}$Center for Gravitational Physics and Quantum Information, Yukawa Institute for Theoretical Physics, Kyoto University, Kyoto 606-8502, Japan}
\affiliation{$^{2}$Departament de F{\'\i}sica Qu\`antica i Astrof\'{\i}sica, Institut de
Ci\`encies del Cosmos,
 Universitat de
Barcelona, Mart\'{\i} i Franqu\`es 1, E-08028 Barcelona, Spain}
\affiliation{$^{3}$Kobayashi-Maskawa Institute, Nagoya University, Nagoya 464-8602, Japan} 
\affiliation{$^{4}$Department of Mathematics, Nagoya University, Nagoya 464-8602, Japan}
\affiliation{$^{5}$Division of Science, School of Science and Engineering, Tokyo Denki University, Saitama 350-0394, Japan}
\affiliation{$^{6}$Department of Physics, Osaka Metropolitan University, Osaka 558-8585, Japan}
\affiliation{$^{7}$Nambu Yoichiro Institute of Theoretical and Experimental Physics (NITEP), Osaka Metropolitan University, Osaka 558-8585, Japan}

\begin{abstract}
\begin{center}
{\bf Abstract}
\end{center}
\noindent 
In our previous papers ~\cite{Amo:2021gcn,Amo:2021rxr,Amo:2022tcg}, we analyzed the asymptotic behavior of future directed null 
geodesics near future null infinity and then we showed a proposition on the accessibility of the null geodesics to future null infinity 
in a specific class of asymptotically flat spacetimes. In this paper, we adopt the retarded time of the Bondi coordinate as the parameter 
for the null geodesics and then see that one can relax the assumptions imposed in our previous studies. 
As a consequence, we obtain a new null-access theorem for generic asymptotically flat spacetimes.  
\end{abstract}

\maketitle

%
%
%

\section{Introduction}

Black holes are characterized by such strong gravity that photons cannot escape from them.
Observation of photon emissions from the neighborhood of a black hole shows us a dark region called the shadow, 
which is reported by Event Horizon Telescope Collaboration~\cite{Akiyama:2019cqa,EventHorizonTelescope:2022wkp}. In the mathematical 
formulation for observation of strong gravity region such as black hole shadow \cite{Amo2023}, asymptotic behavior of null geodesics near 
future null infinity is important because a distant observer is approximately located at future null infinity. 
In our previous papers~\cite{Amo:2021gcn,Amo:2021rxr,Amo:2022tcg}, we have addressed this issue. 

Naively, any null geodesic emanating from near future null infinity in a non-inward direction would trivially reach future null infinity. 
However, this turned out to be rather non-trivial. In Refs.~\cite{Amo:2021gcn,Amo:2021rxr,Amo:2022tcg} 
(see~\cite{Amo:2022tcg} and its erratum for the strongest evaluation so far), it was shown that, in four dimensions, gravity 
affects the null geodesic motion at the leading order in the radial coordinate expansion near future null infinity, while it does not 
in higher dimensions. In particular, sufficient conditions for null geodesics to reach future null infinity were presented. 
This condition constrains both the metric and the initial direction of the null geodesic. 
For the metric, it was assumed that near future null infinity, the null energy condition holds, and the gravitational wave and 
matter radiation are not strong enough compared to the Planck luminosity density~\cite{dyson}. 
For the constraints on the null geodesic, it was assumed that a corresponding photon is emitted in an inward direction at a small angle to a constant 
radial surface or in an outward direction so that the radial coordinate expansion works throughout the geodesic we consider. 

In this paper, we reexamine the analyses in our previous papers, especially in Refs.~\cite{Amo:2021gcn, Amo:2022tcg} and then we 
will relax the assumption. As a consequence, we could have a proposition on accessibility of null geodesics to future null infinity 
which is applicable to the generic four-dimensional asymptotically flat spacetime. We call this theorem the ``{\it null-access theorem},''
which would give fairly optimal conditions that guarantee the accessibility of null geodesics to future null infinity for general situations. 
We will only discuss the four-dimensional case because, in the higher-dimensional case, it has been already shown in Ref.~\cite{Amo:2022tcg, Amo:2021gcn} 
that the null energy condition and the assumptions for the metric  are not required. 

The rest of this paper is organized as follows.
In Sec.~\ref{sec:geo}, we give the radial component of null geodesic equation near future null infinity, 
and present the main proposition (the null-access theorem). 
In Sec.~\ref{sec:proof}, we prove the main proposition. 
Section \ref{sec:summary} is devoted to a summary and discussion. 
In Appendix~\ref{sec:detailed_nullgeo}, we provide a detailed analysis on the null geodesic equations. 
In Appendix~\ref{sec:diffder}, we show that the difference between the total derivative and partial derivative of the position of a photon with respect
to the retarded time along the geodesic is negligible at the leading order in our estimation. 
In Appendix~\ref{sec:u'=0}, we discuss the exceptional case that our main proposition is not directly applicable, but one can discuss the 
accessibility of the null geodesics to future null infinity in a merely simple proof.
In Appendix~\ref{sec:infiniter}, 
the details of the proof of showing the divergence of the radial coordinate 
along the null geodesic is presented by 
studying the null geodesic equations explicitly, while the essential point is given in the main text.
Throughout this paper, 
we assume the metric to be $C^{2-}$ functions ({\it i.e.}, class $C^{1,1}$).

%
%
%

\section{Asymptotics and main proposition}
\label{sec:geo}

In this section, we first give the asymptotic form of the metric and the radial component 
of null geodesic equation in four-dimensional asymptotically flat spacetimes. 
Then, we present our main proposition. The proof will be given in Sec.~\ref{sec:proof}. 

\subsection{Null geodesic equations near future null infinity}

We consider a four-dimensional asymptotically flat spacetime. 
In the Bondi coordinates $\lbrace u, r, x^I \rbrace$, where 
$u$, $r$, and $x^I$ are the retarded time, radial and angular coordinates, respectively, 
the nonzero components of the metric $g_{\mu\nu}$ near future null infinity behave as~\cite{Bondi,Sachs} 
\begin{eqnarray}
g_{uu}=-1+mr^{-1} +\Cr{-2}, \quad
g_{ur} = -1 + \Cr{-2},\nonumber\\
g_{IJ} = \omega_{IJ}r^2 + h^{(1)}_{IJ}r + \Cr{0}, 
\quad g_{uI}= \Cr{0}. \quad\label{metric}
\end{eqnarray}
Here $\omega_{IJ}$ is the metric of the unit two-dimensional sphere, and 
future null infinity is described by the limit of $r\to\infty$ while $u$ is kept finite. 
We note, in case, that expansion coefficients, such as $m$ and $h^{(1)}_{IJ}$, are assumed to be bounded, 
which will be used later.

The integration of $m(u,x^I)$ the over solid angle yields the Bondi mass: 
\begin{eqnarray}
    M(u):=\frac{1}{8\pi}\int_{S^2}md\Omega.
   \end{eqnarray}
As the gauge condition, we impose  
\ba
\det \left(g_{IJ}\right)=\det \left(r^2\omega_{IJ}\right),\label{gauge}
\ea
which gives us $\omega^{IJ}h^{(1)}_{IJ}=0$. 

We basically adopt the retarded time $u$ to parameterize the worldline of a photon in this paper. 
  For the angular vector space, then, we introduce the unit vector $e^K$ (with respect to the metric $\omega_{IJ}$) by
  \begin{equation}
  e^K\,:=\,
  \left(\omega_{IJ} \frac{dx^I}{du}\frac{dx^J}{du}\right)^{-1/2}\frac{dx^K}{du}.
  \end{equation}
For any tensor $\alpha_{IJ}(u,{x^I})$ in angular space, it is useful to define a 
function $\alpha\left(u, {x^I};dx^J/du\right)$ as 
\ba
  \alpha\left(u, {x^I};\frac{dx^J}{du}\right)\,:=\,
  \alpha_{KL}e^Ke^L
 \,=\,  \left(\omega_{IJ} \frac{dx^I}{du}\frac{dx^J}{du}\right)^{-1}\alpha_{KL} \left(u, {x^M}\right) \frac{dx^K}{du} \frac{dx^L}{du}.\label{defalpha}
\ea
Note that $\alpha\left(u, {x^I};dx^J/du\right)$ depends on the direction of $dx^J/du$ but not on its norm.
A tensor 
\ba
\Omega_{IJ}:=\omega_{IJ}  - \frac12 \frac{\partial h^{(1)}_{IJ}}{\partial u} + \frac12 \frac{\partial m}{\partial u} \omega_{IJ},\label{OmagaIJ}
\ea
appears as an important quantity to 
determine the behavior of null geodesics near future null infinity, that was shown in
our previous analysis of Refs.~\cite{Amo:2021gcn,Amo:2021rxr,Amo:2022tcg}.
For $\Omega_{IJ}$, a function defined in Eq.~\eqref{defalpha} is
\ba
\Omega\left(u, {x^I};\frac{dx^J}{du}\right)&=&1-\frac{1}{2}\frac{\partial h^{(1)}}{\partial u}\left(u, {x^I};\frac{dx^J}{du}\right)+\frac{1}{2}\frac{\partial m}{\partial u}\left(u, {x^I}\right). \label{Omega_scalar}
\ea

For a comparison, we summarize the proposition shown in Ref.~\cite{Amo:2022tcg} (See the erratum too. There are crucial corrections): 
\begin{proposition}
  \label{prop0}
Consider a four-dimensional asymptotically flat spacetime in which the metric near future null infinity is written as Eq.~\eqref{metric} 
with the Bondi coordinates by $C^{2-}$ functions.
Suppose that $\Omega_{IJ}$ defined by Eq. (\ref{OmagaIJ}) is positive definite and $\partial m /\partial u \le 0$ holds everywhere 
near future null infinity. 
We define $\Omega_{i}$ as the infimum of $\Omega$, where $\Omega$ is introduced in Eq.~\eqref{Omega_scalar}. 
Then, take a point p with a sufficiently large coordinate value $r=r_0$.
Any null geodesic emanating from $p$ reaches future null infinity if 
\ba
0<\left(\left.\frac{dr}{du}\right|_p-\beta_{\rm crit}\right)^{-1}&=&\Crs{}
\ea
holds, where~\footnote{Note that $\beta_{\rm crit}$ in Ref.~\cite{Amo:2022tcg} represents the same quantity as $\dot r_{\rm crit}$. 
Throughout this paper, we do not use  `dot' notation to distinguish between total and partial derivatives.}
\ba
  \beta_{\rm crit}:=
  \frac{-3+\sqrt{9-6\Omega_i}}{3}\label{beta_crit}. 
\ea 
\end{proposition}
\vspace{1mm}

To be more specific, the phrase {\it ``a sufficiently large coordinate value $r=r_0$''} means that $r=r_0$ is large enough compared to the coefficients of the $r$-expansion of the metric and their derivatives with respect to $u$ and $x^I$.
We will reexamine the assumption in Proposition~\ref{prop0} by carefully analyzing the integrals of quantities involved in the geodesic equations along the geodesic.
This makes the analysis sharp and then one can have a statement on the accessibility of future directed null geodesics 
to future null infinity which is applicable to the generic asymptotically flat spacetime. 

After long calculations, near future null infinity, we can write down the $r$-component of the geodesic equation as 
(see Appendix~\ref{sec:detailed_nullgeo} for the details)
\ba
\frac{d^2 r}{du^2}
&=&\left[2\left(\frac{dr}{du}\right)^2+\left\{3-\frac{\partial h^{(1)}}{\partial u}\left(u,x^I;\frac{dx^J}{du}\right)\right\}\frac{dr}{du}
+\Omega \right]\Big[r^{-1}+\Cr{-2}\Big]. 
\hspace{3mm}\label{nullgeo2}
\ea

\subsection{Main proposition}

Now we are ready to present our main proposition. Before that, we give a useful Lemma:
\begin{lemma}
  \label{lemma}
  Consider a four-dimensional asymptotically flat spacetime in which the metric near future null infinity is written as Eq.~\eqref{metric} 
  with the gauge condition of Eq.~\eqref{gauge}.
  Let $\Omega_{i}$ denote the infimum of $\Omega$,
\ba
\Omega_i:=\inf_{u, {x^I},dx^J/du} \Omega,
\ea
  where $\Omega$ is given in Eq.~\eqref{Omega_scalar}. 
  Then, $\Omega_i$ should satisfy
  \ba
\Omega_i\le1.\label{Oi1}
\ea
\end{lemma}
For the comparison, we stress that the condition $\partial m/\partial u\le0$ was assumed to show $\Omega_i\le1$ (which 
played an important role to discuss whether photons reach future null infinity) in Ref.~\cite{Amo:2022tcg}, 
whereas we do not assume 
$\partial m/\partial u\le0$ here.

This lemma can be proven as follows. The gauge condition of Eq.~\eqref{gauge}, that is, $\omega^{IJ}h^{(1)}_{IJ}=0$ gives us
\ba
\omega^{IJ}\Omega_{IJ}=2+ \frac{\partial m}{\partial u}. \label{omegam}
\ea
Since $\omega^{IJ}$ is given by
$\omega^{IJ}=e_1^{I}e_1^{J}+e_2^{I}e_2^{J}$
with a pair of unit orthonormal vectors, $e_1^{I}$
and $e_2^{I}$, we have
\ba
\omega^{IJ}\Omega_{IJ} = \min_{dx^J/du} \Omega+\max_{dx^J/du} \Omega
\,\ge\, 2\min_{dx^J/du} \Omega,
\ea
for each $u, x^I$. Then, we see
\ba
\Omega_i\,\le\,1+\frac{1}{2}\inf_{u, {x^I}}\frac{\partial m}{\partial u}.\label{Omega_i}
\ea
Since $m$ is bounded, the infimum of $\partial m/\partial u$ must not be positive even though we do not assume $\partial m /\partial u \le 0$ explicitly. 
Therefore, Eq.~\eqref{Omega_i} gives us Eq.~\eqref{Oi1}, which completes a proof of Lemma~\ref{lemma}. 

The proposition of this paper is summarised as follows:
\begin{proposition}
  \label{prop}
  {\bf (Null-access theorem):}~
Consider a four-dimensional asymptotically flat spacetime in which the metric near future null infinity is written as Eq.~\eqref{metric} 
with the Bondi coordinates by $C^{2-}$ functions.
We assume the gauge condition of Eq.~\eqref{gauge}.
We define $\Omega_{i}$ as the infimum of $\Omega$, where $\Omega$ is introduced in Eq.~\eqref{Omega_scalar}. 
Then, take a point p with a sufficiently large coordinate value $r=r_0$.
Any null geodesic emanating from $p$ reaches future null infinity if 
\ba
0<\left(\left.\frac{dr}{du}\right|_p-\beta_{\rm crit}\right)^{-1}&=&\Crs{}\label{con1}
\ea
holds, where, for $\Omega_i>0$, 
\ba
  \beta_{\rm crit}:=
  \frac{-3+\sqrt{9-6\Omega_i}}{3}\label{beta_crit}
\ea 
and, for $\Omega_i\le0$, $\beta_{\rm crit}$ is an arbitrary positive constant. 
\end{proposition}
\vspace{1mm}

There are several remarks. 
(i) First, we note that $\beta_{\rm crit}$ is real by virtue of Lemma~\ref{lemma}. 
(ii) We would stress again that this proposition does not assume $\partial m/\partial u\le0$ nor 
the positive definiteness of $\Omega_{IJ}$ assumed for Proposition \ref{prop0} shown in Ref.~\cite{Amo:2022tcg}. 
(iii) The condition~\eqref{con1} does not include the case of $u'=0$ at $p$, where the prime denotes the derivative with 
respect to the affine parameter. However, we can show that the 
future directed null geodesics with $u'=0$ will reach future null infinity.
See Appendix~\ref{sec:u'=0} for the details. 

%
%
%
\section{Proof of Proposition 2}
\label{sec:proof}

The proof of Proposition \ref{prop} is composed of three steps. 
At the first step, it is shown that $dr/du$ will become nonnegative even for the case with initially negative $dr/du$. 
The second and third steps show that $u$ is kept finite, and that $r$ goes to infinity, respectively.

\subsection{Asymptotic behavior of $dr/du$}
\label{sec:non-}

We first discuss the case with $dr/du<0$ at the initial point $p$, which gives us $\Omega_i>0$ from the assumption of the proposition. 
As was derived in Ref.~\cite{Amo:2022tcg} and its erratum,
there exists $u_1$, which is larger than the value of $u$ at the initial point $p$, satisfying both~\footnote{In erratum of Ref.~\cite{Amo:2022tcg}, $r(u)$ was evaluated as 
\ba
  r(u) &\geq&\frac{1}{2}r_0\left(3\dot{r}(0)^2+6\dot{r}(0)+2\Omega_i\right)\left(2\dot{r}(0)^2+3\dot{r}(0)+\Omega_i\right)^{-1}\left[1+\Crz{-1}\right],\nonumber
\ea
where dot denotes the derivative with respect to $u$. Eq.~\eqref{r(u1)>C_1} is obtained by noting $1+\Crz{-1}>2/3$ for sufficiently large $r_0$.
}
\ba
     \frac{dr}{du}(u_1)&>&0,\\
     r(u_1) &\ge &C_1r_0,\label{r(u1)>C_1}
      \ea
     where $C_1$ is a positive constant defined as
     \ba
     C_1&:=&\frac{1}{3}\left[3 \left(\frac{dr}{du}(0) \right)^2+6\frac{dr}{du}(0)+2\Omega_i \right]\left[2\left(\frac{dr}{du}(0) \right)^2+3\frac{dr}{du}(0)+\Omega_i \right]^{-1}.\label{def:C_1}
\ea
Note that $\beta_{\rm crit}$ defined by Eq.~\eqref{beta_crit} is the largest solution to the quadratic equation for $(dr/du)(0)$ such that 
the numerator in the expression \eqref{def:C_1} of $C_1$ vanishes, but the denominator does not. 
Thus, the condition \eqref{con1} tells us $C_1=1/\Crs{}$ and then Eq.~\eqref{r(u1)>C_1} enables the $r$-expansion adopted here
to work appropriately 
(see Ref.~\cite{Amo:2022tcg} for details). 
For $\Omega_i\le0$, at the initial point $p$ in Proposition \ref{prop}, we have $dr/du|_p>0$ under the condition of Eq.~\eqref{con1}. 
Here we take $r(u_1)=r_0$ at $p$. 

Now, we see $(dr/du)(u_1)>0$ for both $\Omega_i>0$ and $\Omega_i\le0$ cases. 
From now on, in order to deal with both cases in a unified manner with the common notation, we introduce $C_1$, 
which was defined for $\Omega_i>0$ as Eq.~\eqref{def:C_1}, to be $C_1=1$ for $\Omega_i\le0$ . 
Then, we see that the equality holds with in Eq.~\eqref{r(u1)>C_1} for $\Omega_i\le0$ case.

\subsection{Finiteness of $u$}
\label{sec:finiteu}

Let us show that $u$ does not diverge along future directed null geodesics in the current setup. 
Using $h^{(1)}$ for $h^{(1)}_{IJ}$ defined through Eq.~\eqref{defalpha}, 
we define $\beta_1(u, {x^I}, dx^J/du)$ as 
\ba
\beta_1\left(u, {x^I}, \frac{dx^J}{du}\right):=-2\left(3-\frac{\partial h^{(1)}}{\partial u}\right)+2\sqrt{\left(3-\frac{\partial h^{(1)}}{\partial u}\right)^2-\Omega}
\ea
if $u$, ${x^I}$, and $dx^J/du$ satisfy
\ba
\left(3-\frac{\partial h^{(1)}}{\partial u}\right)^2&-&\Omega\ge0,\label{beta1case}
\ea
and, otherwise, we set 
\ba
\beta_1=0.
\ea
If $dr/du>\beta_1$, we see that 
\ba
\frac{1}{4}\left(\frac{dr}{du}\right)^2+\left(3-\frac{\partial h^{(1)}}{\partial u}\right)\frac{dr}{du}+\Omega>0\label{positive12}
\ea
holds, which we will use for the estimate of the left-hand side of Eq.~\eqref{nullgeo2} later. 
Defining $\beta_2$ as  
\ba
\beta_2&:=&\sup_{u, {x^I}, dx^J/du}\beta_1\left(u, {x^I}, \frac{dx^J}{du}\right), \label{dotr2}
\ea
we see that Eq.~\eqref{positive12} holds for $dr/du>\beta_2$. 
But there is still a possibility that $0<(dr/du)( u_1)\le\beta_2$. 
Therefore, let us show by contradiction that, even if we start with $0<(dr/du)( u_1)\le\beta_2$, there exists a constant $u_2>u_1$ , such that $(dr/du)(u_2)>\beta_2$ is satisfied. 
Hence, we impose the condition $dr/du \le \beta_2$ for any $u\ge u_1$ which leads us the contradiction as shown in the next paragraph. 

We first show that $(dr/du)( u)$ is positive for $u>u_1$ by contradiction. 
So let us suppose this positivity is violated for some $u>u_1$, under the assumption $dr/du\le \beta_2$. 
We define the minimum $u_{{\rm min}}$ of $u>u_1$ satisfying $(dr/du)(u)= 0$.
Note that for $u_1<u<u_{{\rm min}}$, $0< dr/du \le \beta_2 $ is satisfied and thus Eq.~\eqref{r(u1)>C_1} gives
\ba
r(u) \ge r(u_1) \ge C_1r_0 . \label{>C_1}
\ea
For $\Omega_i>0$  at $u=u_{{\rm min}}$, Eq.~\eqref{nullgeo2} gives $d^2r/du^2>0$, whereas, since  $(dr/du)( u)$ is positive for $u_1<u<u_{{\rm min}}$, $d^2r/du^2$ should be negative. 
This results in a contradiction.
For $\Omega_i\le0$, more detailed analysis is required.
For $u_1<u<u_{{\rm min}}$, Eq.~\eqref{nullgeo2} together with Eq.~\eqref{dotr_h} gives us~\footnote{Note that the partial derivative with respect to $u$ does not appear. We had some careful estimation. 
See Appendix~\ref{sec:diffder} for the details. }
\ba
\frac{d}{du} \left( r \frac{dr}{du} \right)
&>&\left[-\frac{d}{du}\left(\frac{dr}{du}h^{(1)}\right)+
\hat \Omega 
\right]\Big{[}1+\Crz{-1}\Big{]},\hspace{8mm}\label{dotrdotr}
\ea
where 
\ba
\hat \Omega :=1-\frac{1}{2}\frac{dh^{(1)}}{du}+\frac{1}{2}\frac{dm}{du}.
\ea
Then, the integration of Eq.~\eqref{dotrdotr} implies us 
\ba
r(u)\frac{dr}{du}(u)-r(u_1)\frac{dr}{du}(u_1)
&>&\Big{[}-\frac{dr}{du}(u)h^{(1)}(u)+\frac{dr}{du}(u_1)h^{(1)}(u_1)+u-u_1\nonumber\\
&&\hspace{4mm}-\frac{1}{2}h^{(1)}(u)+\frac{1}{2}h^{(1)}(u_1)+\frac{1}{2}m(u)-\frac{1}{2}m(u_1)
\Big{]}\Big{[}1+\Crz{-1}\Big{]}\nonumber\\
&=&\Big{[}1+\Crz{-1}\Big{]}(u-u_1)+\Crz{0}, \label{rdotr}
\ea
where, in the last equality, we used the condition $0< dr/du \le \beta_2$ and the fact that $\beta_2$, $h^{(1)}(u)$, and $m(u)$ are quantities of $\Crz{0}$.
Recalling $u_1=u_0$ for $\Omega_i\le0$, we obtain $(dr/du)( u_1)>\beta_{\rm crit}>0$ due to Eq.~\eqref{con1}.
Then, Eq.~\eqref{rdotr} gives 
\ba
r(u_{{\rm min}})\frac{dr}{du}(u_{{\rm min}})>r(u_1)\frac{dr}{du}(u_1)+\Crz{0}\,>\,0,
\ea
where in the second inequality,
we used the fact that since the first term is comparable to $r_0^{~1}$,
it dominates over the second term of $\Crz{0}$. This result 
contradicts the definition of $u_{{\rm min}}$.
Thus, $(dr/du)( u)$ is positive for $u>u_1$.

In a similar way to the derivation of Eq.~\eqref{dotrdotr}, we have 
\ba
\frac{dr}{du}(u)-\frac{dr}{du}(u_1)
>\Big{[}r_0^{-1}+\Crz{-2}\Big{]}(u-u_1)+\Crz{-1}. \label{dotr}
\ea
The left-hand side of the above can be arbitrarily large by taking sufficiently large $u$, which means that $(dr/du)(u)>\beta_2$ holds at finite $u$. 

Next, let us investigate the behavior after $(dr/du)( u)>\beta_2$ is achieved.
We set a value of $u\ge u_1$ satisfying $(dr/du)( u)>\beta_2$ as $u_2$. We can take $u_2=u_1$ for the case $(dr/du)( u_1)>\beta_2$. 
Using Eq.~\eqref{positive12}, Eq.~\eqref{nullgeo2} implies 
\ba
\frac{d^2 r}{du^2}&>&\frac{7}{4}\left(\frac{dr}{du}\right)^2r^{-1}\left[1+\Cr{-1}\right]\nonumber\\
&>&\frac{3}{2}\left(\frac{dr}{du}\right)^2r^{-1}\label{fracddotr2}
\ea
for $u$ satisfying $(dr/du)( u)>\beta_2$. 
This means that once $(dr/du)( u)>\beta_2$ is achieved, $(dr/du)(u)$ keeps increasing. 
Therefore,  $(dr/du)( u)>\beta_2$ is satisfied for $u>u_2$. 

From Eq.~\eqref{fracddotr2}, we see 
\ba
\frac{d^2}{du^2}\left[r^{-1/2}(u)\right]<0.\label{d2}
\ea
Integrating this inequality, we have
\ba
\frac{dr}{du}(u)\,>\, C_2r^{3/2}(u)
\ea
for a positive constant
$C_2 \,:=\, r(u_2)^{-3/2}(dr/du)(u_2)$,
which is independent of $u$.
This is rewritten as
\ba
0\,<\,\frac{du}{dr}(u)\,<\, (1/C_2)r^{-3/2}(u).\label{dotr0}
\ea
Integrating this inequality  for $[u_2,u]$, we obtain
\ba
u\,<\,& u_2+(2/C_2)\left[r^{-1/2}(u_2)-r^{-1/2}(u)\right]
\,<\,u_2+(2/C_2)r^{-1/2}(u_2),\label{ufin}
\ea
where we used $C_2>0$ in the second inequality.
This means that for some $u_3$, $u$ is bounded as $u<u_3$.

We now show that the null geodesic actually arrives at future null infinity. 
Suppose, for the sake of contradiction, that the null geodesic stays within the region
$r<r_3$ with some finite constant $r_3$. Since $dr/du>\beta_2$, 
the null geodesic exists within the region $r_2\le r<r_3$. 
Here, we note that every $r$-constant surface crosses $u=u_3$, and furthermore, the region $r_2\le r<r_3$ 
and $u_2\le u<u_3$ is finite. 
Then, any causal curve starting 
from $(u,r)=(u_2,r_2)$ and staying within $r_2\le r<r_3$ inevitably arrives at $u=u_3$.\footnote{We tacitly suppose that, near the asymptotic region, the affine parameter of any future null geodesics
 never become infinite within $r_2\le r <r_3$ and $u_2 \le u<u_3$.
Even without this implicit assumption, the divergence of $r$ can be proven, as shown in Appendix~\ref{sec:infiniter}.} 
This contradicts the property $u<u_3$ that has been proven above. 
Therefore, the value of $r$ of the null geodesic under consideration must diverge. 
In Appendix~\ref{sec:infiniter}, the proof of showing the divergence of $r$
is explicitly presented by studying the null geodesic equations 
for completeness.
This completes the proof of the null-access theorem.

%
%
%

\section{Summary and discussion}
\label{sec:summary}

In this paper, we have
established the null-access theorem (Proposition~\ref{prop}) that 
shows the accessibility of null geodesics to future null infinity in the generic four-dimensional asymptotically flat spacetime. 
The effect of the tiny difference from the exactly flat Minkowski spacetime on the null geodesics is of the same order as the centrifugal force near future null infinity, 
even though the difference of geometry decays as one approaches infinity.
This makes the behavior of null geodesics nontrivial, as we have shown in Refs.~\cite{Amo:2021gcn,Amo:2021rxr,Amo:2022tcg}. 
We proved here that 
$\beta_{\rm crit}$ introduced by Eq.~\eqref{beta_crit}, gives us the minimum initial value of $dr/du$ to guarantee that 
the geodesic will reach future null infinity. 
Note that the condition for the initial direction in our null-access theorem in the present paper is specific to four dimensions.
In higher-dimensional asymptotically flat spacetimes, the effects of the difference from the exactly flat spacetime on 
geodesic equations are of higher order compared to that of the centrifugal force, where $\Omega_i$ is replaced by $1$ as seen in Ref.~\cite{Amo:2021gcn,Amo:2021rxr,Amo:2022tcg}.

Proposition~\ref{prop} gives us a sufficient condition for null geodesics emanating from near future null infinity, not a necessary condition.
In the Vaidya spacetime, null geodesics emitted inwardly at larger angles to $r$-constant surface than those constrained by Eq.~\eqref{con1} 
also reach future null infinity~\cite{Amo:2022tcg}. In this case, null geodesics may pass rather small $r$ regions where expansion with $1/r$ 
does not work. This is why we eliminated such cases, and constrained $dr/du$ as Eq.~\eqref{con1} in Proposition~\ref{prop}. 

We have used the asymptotic behavior of the metric Eq.~\eqref{metric} near future null infinity which is suitable for general relativity. 
However, it would be possible to extend Proposition~\ref{prop} to other gravitational theories 
(see Ref.~\cite{Cao:2021mwx} for an extension of Ref.~\cite{Amo:2021gcn} to Brans-Dicke theory).
This issue is left for future work. 

%
%

\acknowledgments

M. A. is grateful to R. Emparan, S. Mukohyama and T. Tanaka for continuous encouragements and useful suggestions. 
M. A. is supported by the ANRI Fellowship, JSPS Overseas Challenge Program for Young Researchers and Grant-in-Aid for 
JSPS Fellows No. 22J20147 and 22KJ1933. 
M. A., K. I.  and T. S. are supported by Grant-Aid for Scientific Research from Ministry of Education, Science, Sports 
and Culture of Japan (JP21H05182). K. I., H. Y. and T. S. are also supported by JSPS(No. JP21H05189). 
K.~I. is also supported by JSPS Grants-in-Aid for Scientific Research (B) (JP20H01902) 
and JSPS Bilateral Joint Research Projects (JSPS-DST collaboration) (JPJSBP120227705). 
T. S. is also supported by JSPS Grants-in-Aid for Scientific Research (C) (JP21K03551). 
H. Y. is in part supported by JSPS KAKENHI Grant Numbers JP22H01220, 
and is partly supported by Osaka Central Advanced Mathematical Institute 
(MEXT Joint Usage/Research Center on Mathematics and Theoretical Physics JPMXP0619217849).

%
%
%

\appendix
\section{Details of null geodesic equations near future null infinity}
\label{sec:detailed_nullgeo}

In this section, we derive the evolution equation of $r(u)$ by using the geodesic equations and the condition for the geodesic to be null.
See Appendix~\ref{sec:u'=0} for the case of $u'=0$.
Let us define $\left|\left(x^I\right)^\prime \right|$ as
\begin{eqnarray}
    \left|\left(x^I\right)^\prime \right| := \sqrt{\omega_{IJ} \left(x^I\right)^\prime  \left(x^J\right)^\prime }.
    \end{eqnarray}
Near future null infinity, the $r$-component of the geodesic equation is written as
\begin{eqnarray}
r'' &=&-\Gamma^{r}_{uu}{u^\prime}^2 - 2\Gamma^{r}_{ur}u'r' - \Gamma^{r}_{rr} {r^\prime}^2 - 2\Gamma^{r}_{uI} u' \left(x^I\right)^\prime  - 2\Gamma^{r}_{rI} r' \left(x^I\right)^\prime  - \Gamma^{r}_{IJ} \left(x^I\right)^\prime \left(x^J\right)^\prime  \nonumber\\
&=& \left[ \frac12 \frac{\partial m}{\partial u}r^{-1} +  \Cr{-2} \right] {u^\prime}^2
+\Cr{-2}u'r'+\Cr{-3}{r^\prime}^2
+\Cr{-1}u' \left(x^I\right)^\prime   \nonumber\\
&&\hspace{12mm}+\Cr{-1}r' \left(x^I\right)^\prime +
\left[ \left(\omega_{IJ}  - \frac12 \frac{\partial h^{(1)}_{IJ}}{\partial u}\right) r  + \Cr{0} \right] \left(x^I\right)^\prime \left(x^J\right)^\prime \nonumber\\
&=& \left[ \frac12 \frac{\partial m}{\partial u}r^{-1} +  \Cr{-2} \right] {u^\prime}^2+\Cr{-2}{r^\prime}^2 +
\left[ \left(\omega_{IJ}  - \frac12 \frac{\partial h^{(1)}_{IJ}}{\partial u}\right) r  + \Cr{0} \right] \left(x^I\right)^\prime \left(x^J\right)^\prime   ,\hspace{3mm} \label{eqr4d}
\end{eqnarray}
where, in the last line, we used the arithmetic-geometric mean inequalities
\ba
|u'r'|&\le& \frac{1}{2}u'^2+\frac{1}{2}r'^2,\\
{\Big |}u'{\Big |}\left|\left(x^I\right)^\prime \right|&\le& \frac{1}{2}r^{-1}u'^2+\frac{1}{2}r\left|\left(x^I\right)^\prime \right|^2,\label{uI}\\
{\Big |}r'{\Big |}\left|\left(x^I\right)^\prime \right|&\le& \frac{1}{2}r^{-1}r'^2+\frac{1}{2}r\left|\left(x^I\right)^\prime \right|^2.\label{rI}
\ea
Similarly, for the $u$ and $x^I$-components, we have
\begin{eqnarray}
{u}'' &=& -\Gamma^{u}_{uu} {u^\prime}^2  -2\Gamma^{u}_{uI} u' \left(x^I\right)^\prime   -\Gamma^{u}_{IJ} \left(x^I\right)^\prime  \left(x^J\right)^\prime \nonumber\\
&=&
\Cr{-2}{u^\prime}^2  +\Cr{-2} u' \left(x^I\right)^\prime   -\left[ \omega_{IJ} r +\Cr{0}\right] \left(x^I\right)^\prime  \left(x^J\right)^\prime \nonumber\\ 
&=&
\Cr{-2}{u^\prime}^2  -\left[ \omega_{IJ} r +\Cr{0}\right] \left(x^I\right)^\prime  \left(x^J\right)^\prime,\label{equ4d}\\
\left(x^I\right)''&=&-\Gamma^{I}_{uu} {u^\prime}^2  -2\Gamma^{I}_{ur} u' r^\prime   -2\Gamma^{I}_{uJ} u'  \left(x^J\right)^\prime  -2\Gamma^{I}_{rJ} r'  \left(x^J\right)^\prime  -\Gamma^{I}_{JK}  \left(x^J\right)^\prime\left(x^K\right)^\prime \nonumber\\
&=&\Cr{-2}{u^\prime}^2  +\Cr{-4}u' r^\prime  +\Cr{-1} u'  \left(x^J\right)^\prime  +\Cr{-1} r' \left|\left(x^I\right)^\prime \right| +\Cr{0} \left|\left(x^I\right)^\prime \right|^2. \hspace{5mm}\label{eqI4d}
\end{eqnarray}

The condition for the geodesic tangent to be null becomes 
\ba
0
&=&\left[-1+\Cr{-1}\right]u'^2+\left[-2+\Cr{-2}\right]u'r'+\left[\omega_{IJ}r^2+\Cr{1}\right]\left(x^I\right)'\left(x^J\right)'
+\Cr{0}u'\left(x^J\right)'\nonumber\\
&=&\left[-1+\Cr{-1}\right]u'^2+\left[-2+\Cr{-2}\right]u'r'+\left[\omega_{IJ}r^2+\Cr{1}\right]\left(x^I\right)'\left(x^J\right)',\hspace{5mm}
\ea
where we used Eq.~\eqref{uI} in the last line. This gives us
\ba
\left|\left(x^I\right)^\prime \right|^2=\left[r^{-2}+\Cr{-3}\right]u'^2+2\left[r^{-2}+\Cr{-3}\right]u'r'.\label{null}
\ea
Thus, for $u'>0$, Eq.~\eqref{null} is rewritten as 
\ba
\left|\frac{dx^I}{du} \right|^2&=&\left[r^{-2}+\Cr{-3}\right]+2\left[r^{-2}+\Cr{-3}\right]\frac{dr}{du}\label{null0}.
\ea
Using Eqs.~\eqref{equ4d} and \eqref{null}, for $u'>0$, Eq.~\eqref{eqr4d} becomes 
\ba
\frac{d^2 r}{du^2}&=&\Omega_{IJ}r\frac{dx^I}{du}\frac{dx^J}{du}+\Cr{0}\left|\frac{dx^I}{du}\right|^2+2 \Big[r^{-1}+\Cr{-2}\Big]\left(\frac{dr}{du}\right)^2\nonumber\\
&&\hspace{10mm}+\left[\left(1-\frac{\partial m}{\partial u}\right)r^{-1}+\Cr{-2}\right]\frac{dr}{du}.\label{ddotr}
\ea
With Eq.~\eqref{null} and the definition of $\Omega$ of Eq.~\eqref{Omega_scalar}, Eq.~\eqref{ddotr} is expressed as Eq.~\eqref{nullgeo2} in the main text. 

Similarly, Eq.~\eqref{eqI4d} is rewritten as 
\ba
\frac{d^2 x^I}{du^2}&=&\Cr{-2}+\Cr{-4}\frac{dr}{du}+\Cr{-1}\frac{dx^I}{du}+\Cr{-1}\frac{dr}{du}\frac{dx^I}{du}+\Cr{0}\left|\frac{dx^I}{du}\right|^2\label{ddotI}
\ea
for $u'>0$. 

%
%
%

\section{Difference between total and partial derivative}
\label{sec:diffder}

In this Appendix, we show that the difference between the total and partial derivative of 
$h^{(1)}$ and $m$ are of higher order, which will be used in the main text. 
For this purpose, we restrict our attention to the case where 
\ba
0<\frac{dr}{du}(u)\le\beta_2\label{dotcrit2}
\ea
and Eq.~\eqref{>C_1} hold.
Let us check the case of $h^{(1)}$ first. We easily see that 
\ba
\frac{d}{du}\left[h^{(1)}\left(u, {x^I};\frac{dx^J}{du}\right)\right]=\frac{\partial h^{(1)}}{\partial u}+\frac{\partial h^{(1)}}{\partial x^K}\frac{dx^K}{du}+\frac{\partial h^{(1)}}{\partial \left(\frac{dx^K}{du}\right)}\frac{d^2 x^K}{du^2}\label{difference}
\ea
holds.
With Eq~\eqref{dotcrit2}, Eq.~\eqref{null0} shows us  
\ba
\left|dx^I/du \right|=\Cr{-1},\label{dIdu}\\
\left|dx^I/du \right|^{-1}=\Cr{1}.\label{dIdu^-1}
\ea 
Getting back to the concrete expression for $h^{(1)}$ following Eq.~\eqref{defalpha}, we can estimate the quantities appeared in the left-hand side of Eq.~\eqref{difference} as 
\ba
\frac{\partial h^{(1)}}{\partial x^K}&=& \frac{\partial h_{MN}^{(1)}}{\partial x^K} e^M e^N =\Cr{0},\label{hk}\\
\frac{\partial h^{(1)}}{\partial \left(\frac{dx^K}{du}\right)}&=&-2\left(\omega_{IJ} \frac{dx^I}{du}\frac{dx^J}{du}\right)^{-2}\omega_{KL}\frac{dx^L}{du} h^{(1)}_{MN}\frac{dx^M}{du}\frac{dx^N}{du}+2\left(\omega_{IJ} \frac{dx^I}{du}\frac{dx^J}{du}\right)^{-1}h^{(1)}_{KL}\frac{dx^L}{du}\hspace{2mm}\nonumber\\
&=&\Cr{1}.\label{dif1}
\ea
In the last equality for both Eqs.~\eqref{hk} and \eqref{dif1}, we used Eqs.~\eqref{dIdu} and \eqref{dIdu^-1}. 

In addition, with the help of Eqs.~\eqref{dotcrit2} and \eqref{dIdu}, Eq.~\eqref{ddotI} tells us 
\ba
\frac{d^2 x^I}{du^2}=\Cr{-2}.\label{ddorxI}
\ea
Then, using Eqs.~\eqref{>C_1}, \eqref{dIdu} and \eqref{hk}--\eqref{ddorxI} in the estimate, 
Eq.~\eqref{difference} gives us 
\ba
\frac{dh^{(1)}}{du}-\frac{\partial h^{(1)}}{\partial u}=\Cr{-1}=\Crz{-1}.\label{dif2}
\ea
In a similar way, we see that
\ba
\frac{dm}{du}-\frac{\partial m}{\partial u}=\Crz{-1}\label{dif3}
\ea
holds.

The following calculation will be used for the derivation of Eq.~\eqref{dotrdotr}. 
\ba
\hspace{-4mm}\frac{d}{du}\left(\frac{dr}{du}h^{(1)}\right)&=&\frac{d^2 r}{du^2}h^{(1)}
+\frac{dr}{du}\left(\frac{\partial h^{(1)}}{\partial u}+\Crz{-1}\right)\nonumber\\
&=&\left[2\left(\frac{dr}{du}\right)^2+\left(3-\frac{\partial h^{(1)}}{\partial u}\right)\frac{dr}{du}+\Omega
\right]
\Big{[}r^{-1}+\Cr{-2}\Big{]}h^{(1)}\hspace{5mm}\nonumber\\
&&\hspace{10mm}+\frac{dr}{du}\left(\frac{\partial h^{(1)}}{\partial u}+\Crz{-1}\right)\nonumber\\
&=&\frac{dr}{du}\frac{\partial h^{(1)}}{\partial u}+\Crz{-1},\label{dotr_h}
\ea
where we used \eqref{dif2} in the first equality, Eq.~\eqref{nullgeo2} in the second equality, 
and Eqs~\eqref{dotcrit2} and \eqref{>C_1} in the last equality. 
%
%
%

\section{Special case of $u'=0$}
\label{sec:u'=0}
Under the same setup without the condition \eqref{con1} in Proposition \ref{prop}, in this Appendix, we will show that 
the null geodesic emanating from a point $p$ reaches future null infinity if $u'=0$ holds at $p$. 
In the Minkowski spacetime, $u'=0$ implies $r'>0$ for a future directed affine parameter. Without loss of generality, 
following this, we can set the affine parameter so that $r'>0$ holds at $p$. 

For any point with $u'=0$, the null condition of Eq.~\eqref{null} implies that $\left(x^I\right)'=0$ holds at this point.
Then, the $u$-component of the geodesic equation of Eq.~\eqref{equ4d} gives us $u''=0$ at this point. Therefore, 
we can see that $u'=0$ holds at any point along the future directed null geodesic, which means that $u$ is 
kept finite when the affine parameter goes to infinity. 
We also see that $\left(x^I\right)'=0$ holds at any point along the null geodesic due to the null condition of Eq.~\eqref{null}. 
Then, at point with $u'=0$, Eq.~\eqref{eqr4d} becomes
$r''=\Cr{-2}r'^2$
and then we see that~\footnote{Note that a quantity of $\Cr{-2}$ in $r''=\Cr{-2}r'^2$ depends on $u$ and $x^I$. 
We take the infimum of this quantity throughout the spacetime such that $C_3$ is a constant throughout the spacetime. }
\ba
r''>-C_3r^{-2}r'^2\label{lowerbound_r''}
\ea
holds for some positive constant $C_3$. Dividing both sides of Eq.~\eqref{lowerbound_r''} with $r^\prime$ and integrating them, we have
\ba
\log r'>\frac{C_3}{r} +C_4>C_4
\ea
for some constant $C_4$, where we used $C_3>0$ in the second inequality.
This gives us \footnote{Equation \eqref{r'>} guarantees that $r'$ will not become nonpositive.}
\ba
r'>e^{C_4}\lambda+C_5\label{r'>}
\ea
for some constant $C_5$, which implies that 
\ba
\lim_{\lambda\to\infty}r=\infty.
\ea
Therefore, the null geodesic reaches future null infinity.

%
%
%
\section{Divergence of $r$}
\label{sec:infiniter}

In this appendix, 
we explicitly show that $r$ will diverge using the geodesic equations
focusing on $u>u_2$.
Due to Eqs.~\eqref{eqr4d} and \eqref{null}, we have
\begin{eqnarray}
    r'' 
    &=& \Cr{-2}{r^\prime}^2 +
    \left[ \Omega  r^{-1}  + \Cr{-2} \right] u'^2
+2\left[ \left(1  - \frac12 \frac{\partial h^{(1)}}{\partial u}\right) r^{-1}  + \Cr{-2} \right]u'r'\nonumber\\
    &>& -C_6r^{-2}{r^\prime}^2 +
   \left(  - \frac12 \frac{\partial h^{(1)}}{\partial u}+\frac12 \frac{\partial m}{\partial u}\right) r^{-1} 
    u'^2-\frac{\partial h^{(1)}}{\partial u}r^{-1}   u'r' \nonumber \\
  & = & 
    \left[-C_6r^{-2}+
    \left(  - \frac12 \frac{\partial h^{(1)}}{\partial u}+\frac12 \frac{\partial m}{\partial u}\right) r^{-1} \left( \frac{du}{dr}\right)^2-\frac{\partial h^{(1)}}{\partial u}r^{-1}   \frac{du}{dr}\right]{r^\prime}^2\nonumber\\
    &>& \left[-C_6-
    \left|  - \frac12 \frac{\partial h^{(1)}}{\partial u}+\frac12 \frac{\partial m}{\partial u}\right| C_2^{-2}r^{-2}-\left|\frac{\partial h^{(1)}}{\partial u}\right|  C_2^{-1}r^{-1/2}\right]r^{-2}{r^\prime}^2, 
    \label{272}
    \end{eqnarray}
where the prime denotes the derivative with respect to the affine parameter $\lambda$, 
$C_6$ is a positive constant independent of $r(\lambda)$ and we used Eq.~\eqref{dotr0} in the last line. 
Then, there exists a positive value $C_7\left(u,x^I,dx^J/du,u_2\right)>0$  such that it does not depend on $r$ and  
\ba
r''&>&-C_7\left(u,x^I,\frac{dx^J}{du},u_2\right)r^{-2}r'^2\label{r''1}
\ea
holds.\footnote{
 $C_7$ depends on $u_2$ since $C_2$ and the minimum value of $r$ depend on $u_2$.} Note that $C_7\left(u,x^I,dx^J/du,u_2\right)$ is not necessarily small enough.
Let $C_8\left(u_2\right)(>0)$ be the supremum of $C_7\left(u,x^I,dx^J/du,u_2\right)$ for $u>u_2$.
From Eq.~\eqref{r''1}, we see 
\ba
r''
&>&-C_8\left(u_2\right)r^{-2}r'^2
\ea
and then it is rearranged to 
\ba
\left(\log r'\right)'>C_8\left(u_2\right)\left(1/r \right)'\label{r''}
\ea
for $r'>0$.  
Here let $\lambda_1$ satisfy $u\left(\lambda_1\right)>u_2$. 
By integrating Eq.~\eqref{r''} for the interval $[\lambda_1,\lambda]$, we obtain
\ba
\log r'\left(\lambda\right)&>&\log r'\left(\lambda_1\right)+C_8\left(u_2\right)\left(r^{-1}\left(\lambda\right)-r^{-1}\left(\lambda_1\right)\right)\nonumber\\
&>&\log r'\left(\lambda_1\right)-C_8\left(u_2\right)/r\left(\lambda_1\right).
\ea
This gives us
\ba
r'(\lambda)>r'\left(\lambda_1\right)e^{-C_8\left(u_2\right)/r\left(\lambda_1\right)}.\label{r'}
\ea
Integration of Eq.~\eqref{r'} for the interval $[\lambda_1,\lambda]$ yields
\ba
r(\lambda)>r(\lambda_1)+r'\left(\lambda_1\right)e^{-C_8\left(u_2\right)/r\left(\lambda_1\right)}\left(\lambda-\lambda_1\right)
\ea
and then we can see
\ba
\lim_{\lambda\to\infty}r=\infty.
\ea
Thus, $r$ goes to infinity along the current null geodesics while $u$ is kept finite.

\end{document}